\documentclass[aps,prd,superscriptaddress,twocolumn,floatfix]{revtex4}
\usepackage{graphicx, subfigure} 
\usepackage{amsmath} 
\usepackage{amsfonts}
\usepackage{amssymb}

\begin{document}

\title{Evidence   for  a   conformal   phase  in   SU(N)  gauge   theories}
\author{A.   Deuzeman}   \affiliation{Centre   for   Theoretical   Physics,
University  of  Groningen,  9747  AG, Netherlands}  \author{M.P.  Lombardo}
\affiliation{INFN-Laboratori Nazionali di Frascati, I-00044, Frascati (RM),
Italy}  \author{E. Pallante}  \affiliation{Centre for  Theoretical Physics,
University of Groningen, 9747 AG, Netherlands}

\date{\today}
\begin{abstract}
\noindent We  discuss the existence of  a conformal phase  in $SU(N)$ gauge
theories in four dimensions. In this lattice study we explore  the model 
 in the bare parameter space, varying the lattice coupling and bare
 mass.
 Simulations  are carried out with three colors
and  twelve flavors  of  dynamical staggered  fermions  in the  fundamental
representation. The  analysis of  the chiral order  parameter and  the mass
spectrum of the theory indicates the restoration of chiral symmetry at zero
temperature  and the  presence of  a  Coulomb-like phase,  
depicting a  scenario compatible with the existence of  an infrared stable 
fixed point at nonzero
coupling.  Our analysis  supports  the  conclusion that  the  onset of  the
conformal window for QCD-like theories is smaller than $N_f=12$, before the
loss of asymptotic  freedom at sixteen and a half  flavors. We discuss open
questions and future directions.
\end{abstract}
 \maketitle

\section{Introduction}
\begin{figure}
\vspace*{0.5 truecm}
\includegraphics[width=8 truecm]{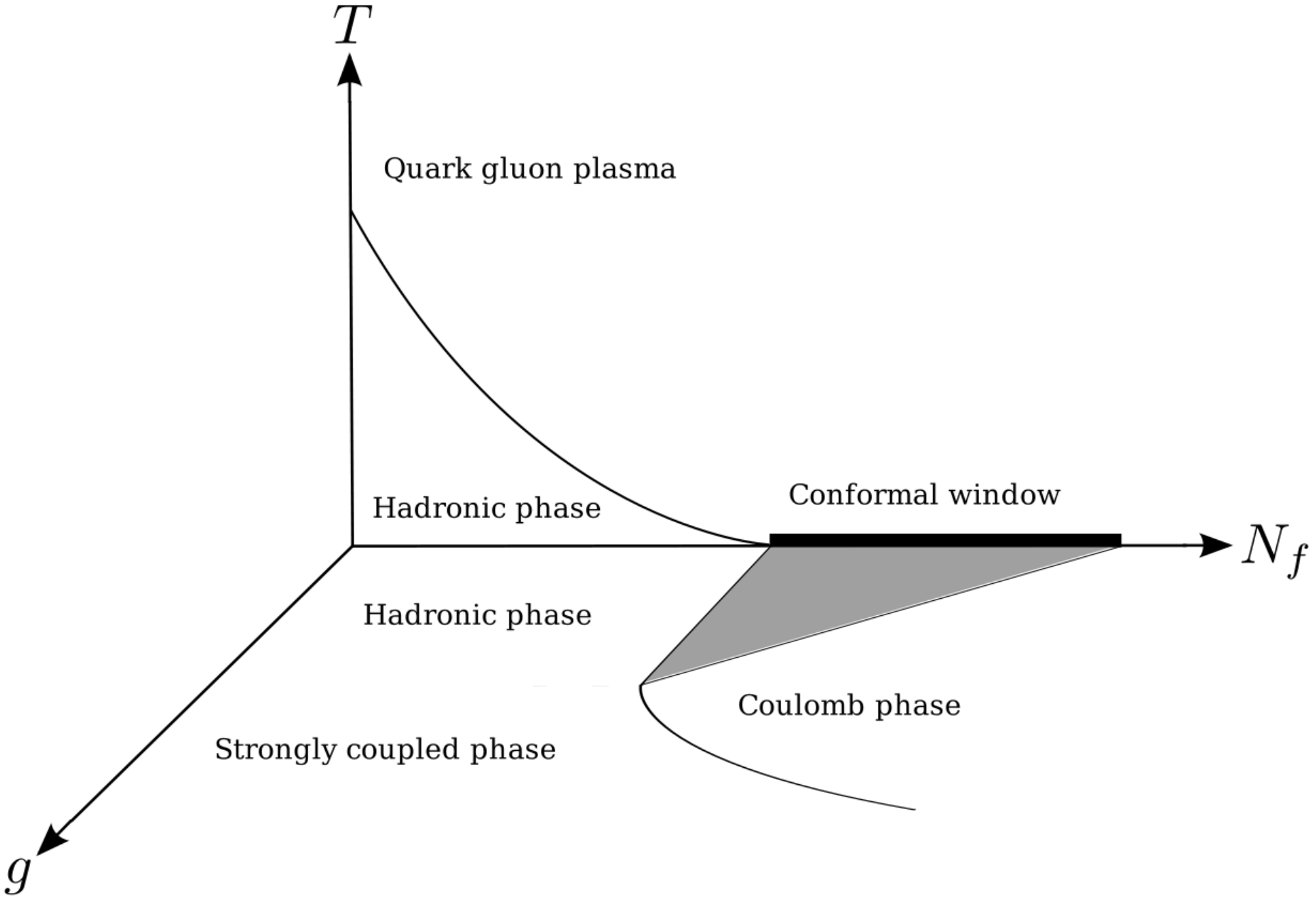}
\caption{\label{Phaseplot}  A  projected  view  of  the  phase  diagram  of
QCD-like theories in the temperature  ($T$), flavor number ($N_f$) and bare
coupling ($g$)  space. In the T-$N_f$  plane, the critical line  is a phase
boundary  between  the chirally  broken  hadronic  phase  and the  chirally
symmetric quark  gluon plasma, the zero  temperature end point  of which is
the onset of the conformal  window. The zero temperature projected plane is
inspired     by     the     scenario    in     Refs.~\cite{appelquist_1996,
miransky_conformal_1997}, see Fig.~\ref{fig:manyphases}.}
\end{figure}

With  the imminent  activity of  the LHC  experiments and  the quest  for a
theory  describing  fundamental  forces  beyond  the  electroweak  symmetry
breaking scale, renewed interest has  arisen in the most elusive aspects of
gauge   theories.   In  particular,   the   possibility   of  an   emergent
quasi-conformal symmetry in theories  with fermionic content has attained a
strong experimental  appeal. There are  multiple reasons for  pursuing this
search. Resolving  conformal behavior  would complete our  understanding of
the phase  diagram of gauge theories  by varying temperature  and number of
flavors, as sketched in Fig.~\ref{Phaseplot}. It sheds light on how the low
temperature and large flavor  number quasi-conformal phase may be connected
to the high temperature and  low flavor number quark-gluon plasma phase. It
is   essential  for   theoretically  establishing   or   excluding  walking
technicolor-type theories and  more generally strongly interacting dynamics
above the electroweak symmetry breaking scale. Finally, elucidating the way
conformal  symmetry  or  its   remnants  drive  the  dynamics  of  particle
interactions with  or without  supersymmetry contributes to  clarifying the
possible  connection of  field theory  to  string theory  that the  AdS/CFT
correspondence seems to imply.

In the early eighties our understanding of the perturbative behavior of non
Abelian   gauge    theories   was   enriched   by    two   seminal   papers
\cite{caswell_asymptotic_1974,banks_phase_1982}.  It  was  noticed  that  a
second zero  of the two-loop  beta function of  an SU(3) gauge  theory with
$N_f$ massless fermions in  the fundamental representation appears for $N_f
\gtrsim 8.05$, at $g^{*2} \ne 0$,  before the loss of asymptotic freedom at
$N^c_f =  16\frac{1}{2}$. This fact  implies, at least  perturbatively, the
appearance of an infrared fixed  point (IRFP). The fixed point moves closer
to zero coupling as the number of flavors approaches $N^c_f$.
The dynamics of chiral symmetry have  led to the discovery of the conformal
window in QCD-like theories~\cite{appelquist_1996,
miransky_conformal_1997}: chiral  symmetry breaking can only  occur below a
critical  number  of  flavors  $N_f^*$.  Between $N_f^*$  and  $N_f^c$  the
conformal window opens up. Finding  the actual value of the critical number
of  flavors $N_f^*$  at which  chiral symmetry  breaking takes  place  is a
non-perturbative problem  for which the  lattice approach is the  method of
choice  \cite{Pallante2009}.  Recent studies  have  provided evidence  that
$N_f=8$       lies      within       the       hadronic      phase       of
QCD~\cite{appelquist_lattice_2008,  deuzeman_physics_2008,  Fleming:2008gy,
appelquist_lattice_2009, Jin:2008rc}.  A recent study of  the SU(3) running
coupling~\cite{appelquist_lattice_2008, appelquist_lattice_2009}  by use of
the  lattice Schr\"odinger  functional has  concluded that  $N_f=12$ should
already  be in  the  conformal window.  Other  numerical studies,  however,
challenged     this    conclusion~\cite{Fodor:2009ff,    Hasenfratz:2009kz,
Jin:2009mc}. This is hardly surprising,  given that $N_f=12$ should be very
close   to  the   critical  number   of  flavors~\cite{appelquist_new_1999,
braun_chiral_2006,   ryttov_conformal_2007,   Sannino:2008ha},   making   a
numerical study particularly delicate.

The      current      strategy,      complementary     to      that      of
Refs.~\cite{appelquist_lattice_2008,  appelquist_lattice_2009}, is inspired
by  the physics  of phase  transitions; it  allows for  the  exploration of
multiple  aspects of the  theory in  different regimes  and regions  of the
phase diagram,  in order to probe  the existence and properties  of an IRFP
inside and outside its basin of attraction.

This paper is organized as follows. In Section \ref{sec:strategy} we review
previous  theoretical work,  in  particular the  scenario for  conformality
originally proposed in Refs.~\cite{appelquist_1996, miransky_conformal_1997},    
and     define    our    strategy.  
Section \ref{sec:simulation} shortly describes the simulations and the 
observables of this work. Section \ref{sec:bulk} presents the results 
on the bulk chiral phase transition, while Section 
\ref{sec:chiral} further  discusses    the   behavior   of    the   chiral   order
parameter. Here, several subsections describe various
theoretically motivated models, and related fits for the mass dependence of the chiral 
condensate.  
Section \ref{sec:spectrum}  addresses the  spectrum, and  it is
organized  in two subsections. The first  one discusses  the interrelation
between the spectrum results and the pattern of chiral symmetry. The second
subsection, similar in  spirit to Ref.~\cite{damgaard_lattice_1997}, argues
that  the  lattice  spacing  increases  when decreasing  the  coupling,  as
expected  of a  negative  beta  function; finally,  it  uses the  numerical
results  combined with  the perturbative  input to  argue in  favor  of the
existence of a  zero of the beta function.  In Section \ref{sec:conclusion}
we summarize the  results, draw our conclusions and  briefly discuss future
directions.

\section{\label{sec:strategy} A scenario for conformality and a lattice strategy}

\begin{figure}
\vspace*{0.5 truecm}
\includegraphics[width=8.0truecm]{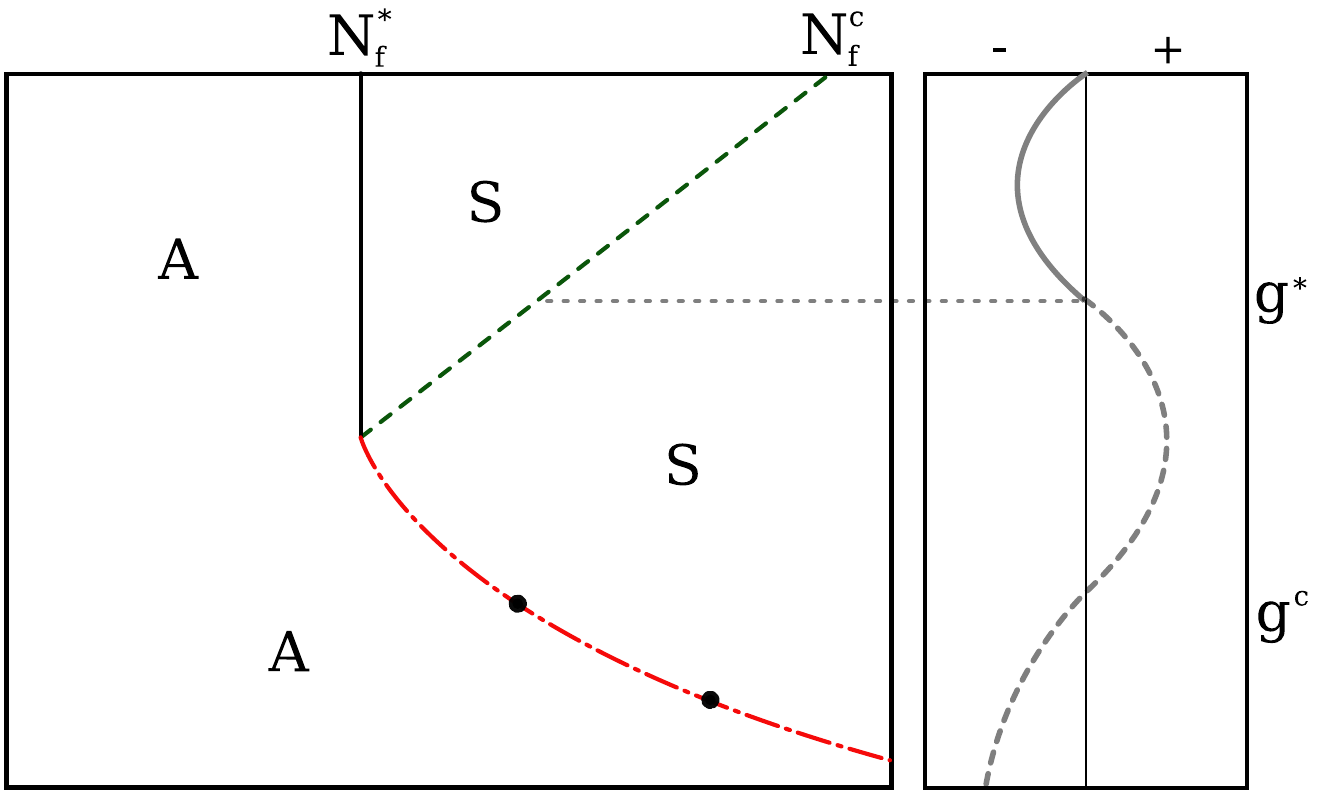}
\caption{\label{fig:manyphases} {(color online) Phase diagram of an $SU(3)$
gauge  theory  with  fundamental  fermions  in the number of 
flavors $N_f$ - bare coupling $g$ plane  after
Ref.~\cite{miransky_conformal_1997}.  Theories   for  $N_f  <   N^*_f$  are
QCD-like  in the  continuum, while  for  $N^*_f <  N_f <  N^c_f$ develop  a
conformal  phase. S  and  A  refer to  chirally  symmetric and  asymmetric,
respectively. The  dashed(green) line qualitatively  indicates the location
of  the Banks-Zaks IRFP  \cite{banks_phase_1982}. The  dot-dashed(red) line
indicates a  lattice bulk transition,  which has been observed  at $N_f=12$
and $N_f=16$.   
The line at $N_f=N_f^*$ represents the conformal phase
transition~\cite{appelquist_1996,miransky_conformal_1997},  which is absent
in the  original Banks-Zaks  scenario. The beta  function on  the conformal
side is also sketched.}}
\end{figure}

The strategy of this study  has received heuristic guidance from the scenario
depicted   in  Refs.~\cite{appelquist_1996,   miransky_conformal_1997}  and
sketched  in  Fig.~\ref{fig:manyphases}.

The zero temperature phase diagram of Fig.~\ref{fig:manyphases}, originally proposed in 
Ref.~\cite{miransky_conformal_1997}, is of course conjectural 
at this stage: even the existence of the conformal window itself needs to be verified by an {\em ab initio} calculation. 
The scenario is based on analytic, necessarily approximate estimates, and
{\em ab initio} lattice studies are also needed to clarify the shape and
nature of the various lines. Importantly, the line of IRFP is
not a phase transition in the scenario of Ref.~\cite{miransky_conformal_1997}, while it is
a chiral transition in the one of Ref.~\cite{banks_phase_1982}, known as the 
Banks-Zaks scenario. The shape of the line of IRFP is of course scheme dependent. The nature of
the phase transitions on each line of Fig.~\ref{fig:manyphases}, in particular on the bulk transition line -- which is relevant in the context of the search for an UVFP at strong coupling -- 
and  the way the lines merge depend on the details of the dynamics. 
This is why it is important to carry out a lattice study.

For our present scope, it suffices to bear in mind that
a conformal window, if any, should be preempted by a zero temperature 
lattice chirally symmetric phase. This is a robust feature, which does
not depend on any of the interesting details of the phase diagram 
outlined above.

We  will  implicitly  assume  the validity of that scenario in the 
lattice bare coupling 
$g$ space, where we shall work in the rest of this paper. 

On the weak coupling side of Fig.~\ref{fig:manyphases}, for any $N_f<N_f^c$, 
the continuum limit exists for $g\to 0$, due to asymptotic freedom.
Should the IRFPs and the conformal window exist, the corresponding lines in 
Fig.~\ref{fig:manyphases} have a mapping onto the phase diagram of the continuum theory.
We add that, if an UVFP at strong coupling \cite{Kaplan:2009kr} exists, the line of bulk transitions signals the emergence of a new continuum limit on the strong coupling side of Fig.~\ref{fig:manyphases}.
The existence of UV fixed points at strong coupling in four dimensions is a long standing problem in field theory. Second order phase transitions at strong coupling
are natural candidates for such fixed points. 
Their non trivial critical dynamics could signal the emergence of an interacting theory, distinct from the asymptotically free dynamics of QCD.

Following Fig.~\ref{fig:manyphases}, at  a given $N_f>N^*_f$ and increasing
the coupling  from $g=0$, one crosses  the conformal line,  location of the
IRFPs, going from  a chirally symmetric (S) and  asymptotically free phase
(quasi-conformal  phase) to a  symmetric, but  not asymptotically  free one
(Coulomb-like or QED-like phase). A phase transition need not be associated with
the  line of  IRFPs, differently  from what  was originally  speculated in
Ref.~\cite{banks_phase_1982}. At  even larger couplings, a  transition to a
strongly coupled  chirally asymmetric  (A) phase will  always occur  in the
lattice  regularized theory.  The latter  is referred  to as  a  bulk phase
transition.  In the  symmetric  phases at  nonzero  coupling the  conformal
symmetry  is  still broken  by  ordinary  perturbative contributions.  They
generate the running of the coupling constant which is different on the two
sides of the symmetric phase. See Ref.~\cite{miransky_conformal_1997} for a
detailed  discussion  of  this  point.  We emphasize  that  in  the  region
considered  in this  paper  conformal  symmetry would  still  be broken  by
Coulombic forces.

A theory in the hadronic phase, $N_f<N_f^*$, has a thermal phase transition
in the  continuum from a  low temperature chirally  broken phase to  a high
temperature chirally  symmetric --  quark gluon plasma  -- phase.  Thus, as
argued in  Ref.~\cite{deuzeman_physics_2008}, the observation  of a thermal
transition in the  continuum limit is incompatible with  the existence of a
conformal  fixed point,  see Fig.~\ref{Phaseplot}.  It is  also  clear from
Fig.~\ref{fig:manyphases} that the presence of  a Coulomb-like phase next to the
bulk  transition at  weaker coupling  is  a distinguishing  feature of  the
conformal phase.  Here,  the  non-perturbative  beta  function  should  be
positive, implying  a weakening of  the effective coupling  over increasing
distances. The appearance  of such a region is,  in principle, a sufficient
condition  for  the existence  of  an  IRFP,  since the  perturbative  beta
function of  $SU(3)$ with $N_f<16\frac{1}{2}$ in the  extreme weak coupling
regime is  known to be negative.  Note, however, that the  beta function is
not universal away from fixed points with diverging correlation lengths and
one can  therefore not  exclude {\em a  priori} the appearance of spurious
fixed     points    at    intermediate     values of the coupling 
constant~\cite{damgaard_lattice_1997}. The reader should keep in mind that 
we will work with a lattice beta function, please see  
Section \ref{sec:betafunc} for a caveat and discussions of this point.  

The evidence  presented here thus consists  of a few  components. First, it
will be demonstrated that the  location of the transition from the chirally
symmetric to the broken phase  is not sensitive to the physical temperature
and  is therefore  compatible with  a  bulk nature.  Subsequently, we  will
present a detailed study of the mass dependence of the chiral condensate on
the weak coupling  side of the bulk transition,  which clearly favors exact
chiral symmetry.  Finally, the behavior of  the mass spectrum  close to the
bulk transition will be studied, and found to be compatible with a positive
beta function, similarly to the     observations of
Ref.~\cite{damgaard_lattice_1997} for $N_f=16$, and the restoration of 
chiral symmetry.   These   results   are
consistent with the scenario for conformality of Fig.~\ref{fig:manyphases}.

\section{\label{sec:simulation} The simulations and the observables}

We have simulated an  $SU(3)$ gauge theory with twelve flavors of
staggered fermions in the  fundamental representation. We used
 a tree level  Symanzik improved gauge action to suppress lattice
artifacts,   and  Kogut-Susskind   (staggered)  fermions   with   the  Naik
improvement scheme,  that effectively  extends the Symanzik  improvement to
the matter content. 

High statistics runs were performed at fixed bare quark mass $am= 0.05$
over an extended range of bare lattice couplings, on $16^3\times8$ and  $16^4$
lattices. At two selected couplings, $6/g_L^2 = 3.9$ and $6/g_L^2 = 4.0$ 
we have performed runs on lattices $20^3 \times 32$, 
$24^4$, $32^4$ and five masses $am= 0.025, 0.04, 0.05, 0.06, 0.07$. 
The thermalization of all runs was extensively verified by monitoring the stability of
averages and uncertainties as a function of the discarded number of sweeps, 
and bin size. In addition, we have verified the decorrelation
from initial conditions by performing simulations with ordered
and random starts for a few selected couplings and masses.

We have measured gauge and fermionic observables including the average
plaquette, the Polyakov loop, the interquark potential,
the chiral condensate and its susceptibility, the meson 
spectrum.  We report here on our results for the chiral condensate and 
the meson spectrum. We
underscore that staggered fermions have  a remnant of exact chiral symmetry
which allows a  precise definition of the chiral order parameter -- 
the condensate $\langle\bar\psi\psi\rangle$ -- also on a coarse lattice.

\section{\label{sec:bulk} The Bulk Transition}

\begin{figure}
\vspace*{0.5 truecm}
\includegraphics[width=8.0truecm]{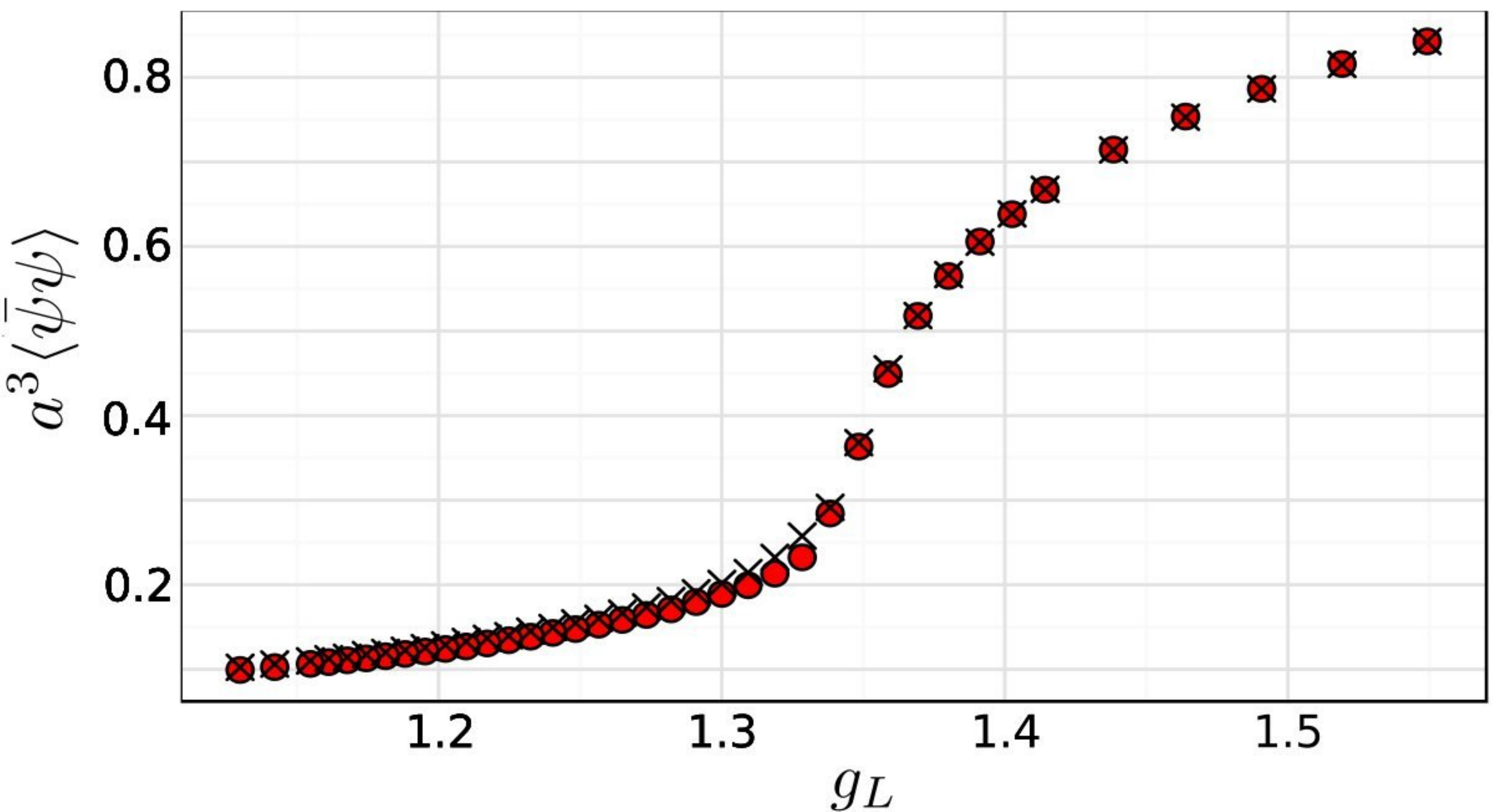}
\caption{\label{fig:bulk} {(color online) The bulk transition in the chiral
condensate  for $am  =0.05$  on lattices  of  $16^3\times8$ (circles),  and
$16^4$ (crosses) as a function of the bare lattice coupling $g_L$. Data are
shown in the  range $6/g_L^2 = 2.5$ to 4.7. The  location of the transition
is identical,  while the curves describe physics  at temperatures differing
by a factor of two. Simulation errors are within symbol size.}}
\end{figure}
Fig.~\ref{fig:bulk} shows our results for the chiral condensate 
at a fixed value of the bare quark mass $am = 0.05$ and for two volumes
 $16^3\times8$ and $16^4$, differing by a factor of two in their temporal extent $N_t$.
 The results display a sudden variation of the chiral order
parameter as  a function of the  bare lattice coupling  constant $g_L$, for both 
$N_t$.  At this point one notices that the temperature of the  system is related to  
the lattice temporal
extent as  $T=1/a(g_L)N_t$, with $a(g_L)$  the lattice spacing for  a given
lattice coupling.  From Fig.~\ref{fig:bulk} one infers  that the phase transition
-- or  rapid crossover  --  happens  at identical  values  of the  critical
coupling $g_L^c  = 1.35(3)$, thus  implying they occur at  vastly different
physical temperatures. Hence, one concludes that the observed transition 
(or crossover) is
driven purely by the bare coupling constant itself and is therefore of bulk
nature. Further  information on this behavior, with a refined scaling study, 
might shed  light on
the occurrence of a conjectured  ultraviolet fixed point at strong coupling
in the continuum theory \cite{Kaplan:2009kr}.

The results of Fig.~\ref{fig:bulk} beg for a detailed analysis of the behavior
of the chiral condensate at weaker couplings, in order to discriminate between a genuine 
phase transition to a chirally symmetric phase, and a rapid crossover
to a phase where chiral symmetry is still broken.

\section{\label{sec:chiral} The chiral condensate at $6/g_L^2 = 3.9$ and $6/g_L^2 = 4.0$}

In order  to be able to extract information  on the symmetry of
the  vacuum - chiral  symmetry broken  or restored  - by  extrapolating the
condensate to  the chiral limit, we  need to measure it  at infinite volume
and at sufficiently light values of the quark masses. Light here means that
the dynamics of  the system is not yet dominated by  the amount of explicit
chiral symmetry breaking. This study, being of course extremely
demanding from the point of view of numerical resources, was performed
for two relevant selected couplings.  We will  first address the issue of
systematic errors, then  we will consider and compare several theoretically motivated
parameterizations, appropriate for chirally broken or symmetric phases. 

\subsection{Aspects of systematics}
\label{sec:systematic}

\begin{figure}
\vspace*{0.5 truecm}
\includegraphics[width=8 truecm]{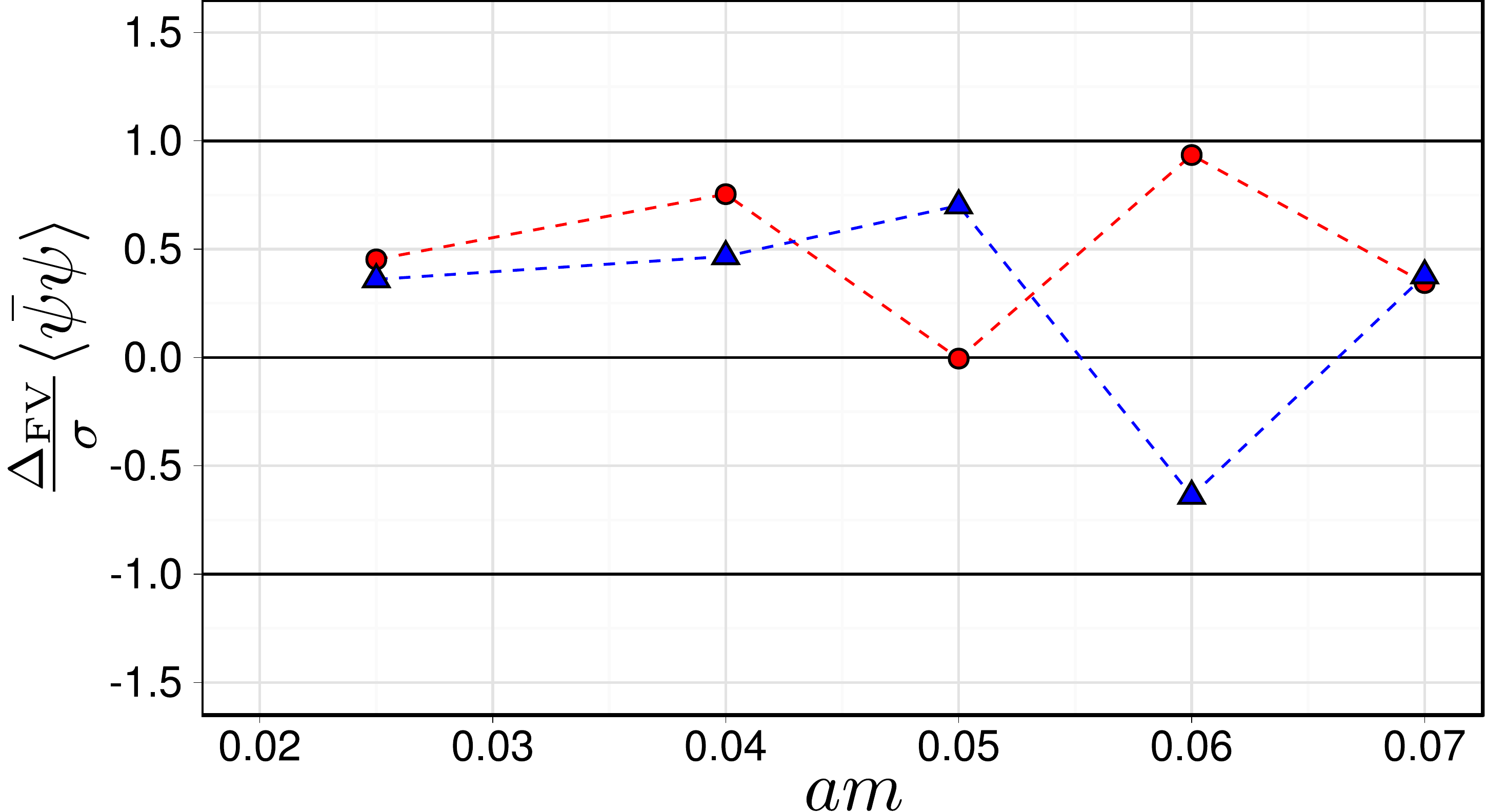}
\caption{\label{fig:fv_effect}  {(color   online)  Observed  finite  volume
effects   in   the  chiral   condensate,   displayed   as  the   difference
$\Delta_\mathrm{FV}$ between the measurements  at the two largest available
volumes ($24^3 \times  24$ and $32^3 \times 32$ for  the lowest mass, $20^3
\times 32$  and $24^3  \times 24$  for the other  masses) divided  by their
combined standard  deviation $\sigma$. Blue triangles  indicate results for
$6/g_L^2=3.9$, red circles those for  $6/g_L^2=4.0$. A value of less than unity
(within the  band) implies that finite  volume effects are  within a single
standard deviation of each other and therefore statistically irrelevant.}}
\end{figure}

To reach the infinite  volume limit within statistical errors, measurements of the chiral 
condensate 
were performed on  three different volumes for each mass,  up to $32^4$ for
the smallest  masses, and  the difference between  the largest  two volumes
found to be  smaller than both the difference  between the smallest volumes
and the statistical uncertainty in all measurements, as can be gleaned from
Fig.~\ref{fig:fv_effect}  and Table \ref{tab:pbp_fvol}.  The data  set used
for the extrapolation to the chiral limit thus consists of the measurements
at  lattice  volumes $24^4$,  which can  be
considered  as  infinite volume  measurements  within  their errors,  again
according to Fig.~\ref{fig:fv_effect} and Table \ref{tab:pbp_fvol}.
\begin{table}
\begin{ruledtabular}
\begin{tabular}{|c|c|c|c|c|}
  $6/g_L^2$ & $am$  & $a^3\langle\bar{\psi}\psi\rangle_\mathrm{N_s = 20}$ &
$a^3\langle\bar{\psi}\psi\rangle_\mathrm{N_s         =        24}$        &
$a^3\langle\bar{\psi}\psi\rangle_\mathrm{N_s = 32}$\\  \hline 3.9 & 0.025 &
0.07638(22)  &  0.07693(07)  &  0.07697(07)  \\ &  0.040  &  0.12107(10)  &
0.12092(17)  & \\  &  0.050 &  0.15018(17) &  0.15018(10)  & \\  & 0.060  &
0.17918(17) &  0.17897(14) & \\  & 0.070 &  0.20776(19) & 0.20768(10)  & \\
\hline 4.0 &  0.025 & 0.07212(13) & 0.07202(10) & 0.07206(05)  \\ & 0.040 &
0.11366(09) & 0.11360(10)  & \\ & 0.050 & 0.14093(19) &  0.14079(07) & \\ &
0.060 & 0.16775(16) & 0.16787(11) &  \\ & 0.070 & 0.19476(11) & 0.19470(13)
& \\
\end{tabular}
\end{ruledtabular}
\caption{\label{tab:pbp_fvol}Comparison  of the measured  chiral condensate
at different  volumes, with  varying bare masses  $am$ and for  two lattice
couplings  $6/g_L^2 =3.9$  and $4.0$.   For all  masses, the  measurements at
volume $N_s  \times  N_t  = 24^3  \times  24$ differ  from the adjacent volumes by  
less than their
statistical uncertainty.  We  therefore use the $24^4$ measurements
as input to the chiral extrapolations of Table~\ref{tab:pbp}.}
\end{table} 

Evidence  that  we  are  considering  sufficiently light  quark  masses  is
provided  by the  mass  dependence of  the  condensate itself,  and by  our
results for the spectrum  in section \ref{sec:spectrum}, where we further
elucidate this  aspect.  

As for the issue of the continuum limit, we remind the  reader that all
the measurements are performed at a  fixed value of the lattice spacing and
no extrapolation to the continuum  limit is considered. 
On the other hand, in the scenario of Fig. 1, there is only one symmetric 
phase at large $N_f$. Hence, once chiral symmetry is restored,
it should stay so till the continuum. A preliminary study towards weak 
coupling has revealed  no sign of further phase transitions, thus confirming
this scenario.   Being notoriously
difficult  to  directly  probe  the  IRFP  with a  lattice  study,  we  are
collecting  precisely  those measurements  at  finite  lattice spacing  and
varying  lattice coupling  that can  provide  a combined  evidence for  the
restoration of  chiral symmetry and for  the existence of  the peculiar non
asymptotically free regime that  precedes the IRFP for decreasing coupling,
a feature proper to non abelian gauge theories with a conformal phase.

\subsection{Fits motivated by a possible Goldstone phase}
\label{sec:goldstone}
The functional forms discussed here would be appropriate if the bulk behavior were not
to be associated to a true chiral transition. For instance, it might just 
be due to a generic rapid crossover, or to a genuinely lattice transition between two 
phases with different ordering. In this case the range of couplings 
between $6/g_L^2 = 3.9$ and $6/g_L^2 = 4.0$  would still belong 
to the phase with broken chiral symmetry. 
We have thus considered the following functional form:
\begin{equation}    
\langle\bar \psi \psi\rangle  = Am + Bm \log (m) + \langle\bar \psi \psi\rangle_0\, ,
\label{eq:goldstone}
\end{equation}
where the parameters were all left free, giving 
fits with two degrees of freedom, or in turn constrained to
zero. The logarithmic mass dependence is typical of a chirally broken phase for a 
QCD-like theory in four dimensions at zero temperature.

The results of the fits to Eq.~\ref{eq:goldstone}
are summarized in Table~\ref{tab:goldstonefit}. 
The linear fits -- case $B=0$ also used in \cite{Jin:2009mc} -- 
produce an intercept different from zero, 
but are highly disfavored by their large $\chi^2$. 
The inclusion of the term $m \log (m)$ considerably improves
the quality of the fits. Those with free intercept  $\langle\bar \psi \psi\rangle_0$
gave an extrapolated value consistent with zero, and in agreement with the fit obtained 
by constraining $\langle\bar \psi \psi\rangle_0 =0$. Both fits are satisfactory,
and imply that 
the chiral condensate in the chiral limit is zero within errors. 
In conclusion, a conventional picture of the Goldstone phase
seems not to be supported by our data.
\begin{table}
\begin{ruledtabular}
\begin{tabular}{|c|c|c|c|c|}
$6/g_L^2 $ & A & B &  $\langle\bar \psi \psi\rangle_0$ & $\sqrt{\chi^2\,\mathrm{dof}}$  \\
\hline
 3.9    & 2.70(3) & -0.103(13) & 0.00013(54) & 0.68  \\
        & 3.12(3)   &  0 (F) & 0.0043(3)   & 3.12  \\  
        & 2.682(5) & -0.107(2) & 0 (F)   & 0.56 \\
\hline
 4.0    & 2.48(2) & -0.120(10) & -0.00091(42) & 0.51  \\
        & 2.73(1) & 0 (F)  & 0.0041(5) & 3.74  \\
        & 2.519(8) & -0.099(3) & 0 (F) & 0.56  \\
\end{tabular}
\end{ruledtabular}
\caption{\label{tab:goldstonefit}
Fits to $\langle\bar \psi \psi\rangle = A m + B m \log m + \langle\bar \psi \psi\rangle_0 $}
\end{table}

\subsection{Fits with an anomalous dimension}

We considered the functional form 
\begin{equation}
\langle\bar \psi \psi\rangle = A m ^{1/\delta} + Bm + \langle\bar \psi \psi\rangle_0\, ,
\label{eq:anomalous}
\end{equation}
 containing an anomalous dimension, whose effect is parameterized by the exponent $\delta$.
Since the fits described in section  \ref{sec:goldstone} already 
suggest that a curvature in the behavior of the chiral condensate as a
function of the mass  is mandatory, we started by setting the linear term  
to zero. We note that analogous fits were used in the past
 to analyze QED in its symmetric phase, close to the strong
coupling transition in Ref.~\cite{Kocic:1990fq},  
even if a more satisfactory account of the
data requires the consideration of the magnetic equation of state,
which is going to be discussed in the next section. 
Results for these fits are reported in Table \ref{tab:anomfit}.
All fits to Eq.~(\ref{eq:anomalous}) with $B=0$ are satisfactory, with a chiral condensate
compatible with zero in the chiral limit. This was checked, as before,
by comparing fits with free intercept, and fits with 
$\langle\bar \psi \psi\rangle_0 = 0$. 

One might still suspect that a fit combining a power-law term and a linear
term, with a non zero intercept might still  
accommodate the data, hence indicating chiral symmetry breaking.
For instance, a linear term can arise because of the additive renormalization
of the chiral condensate -- see e.g. \cite{Ejiri:2009ac} 
for a discussion of this term in the context of the QCD thermal transition.

For completeness we have performed fits to Eq.~(\ref{eq:anomalous}) with the inclusion of 
a linear term. As expected from the near degeneracy between a power law with 
$1/\delta \approx 1$ and a linear term, the uncertainties coming from a Marquardt-Levenberg 
minimization of $\chi^2$ are huge, ranging from 100 \% to 10000 \%. 
In Table \ref{tab:anomfit} we simply quote the central 
results, omitting the errors. Studies able to disentangle the effect of linear
scaling violations \cite{Ejiri:2009ac} were using an exact form for the scaling
function which is not available here.  
In conclusion, the behavior of the fits to Eq.~(\ref{eq:anomalous}) says 
that an additional linear term, or any analytic term in Eq.~(\ref{eq:anomalous}), is 
redundant for our data.  

To acquire a feeling about the possible relevance of
a linear term, we have also performed a sequence of fits, constraining the
exponent to several values in the acceptable range given by
the fit errors. The results are again summarized
in Table~\ref{tab:anomfit}.
It appears that the coefficient of the linear term  smoothly
changes from positive to negative, while the intercept - the chiral
condensate in the chiral limit - remains consistent with zero throughout
at $6/g_L^2=3.9$, and becomes slightly negative at $6/g_L^2=4.0$.
We thus again conclude that our data point at exact chiral symmetry.
\begin{table}
\begin{ruledtabular}
\begin{tabular}{|c|c|c|c|c|c|}
$6/g_L^2 $ & A & $1 / \delta$ & B &  $\langle\bar \psi \psi\rangle_0$ & $\sqrt{\chi^2\,\mathrm{dof}}$  \\
\hline
 3.9    & 3.00  (F)   & 0.960  (F)  & -0.30 (F)   & -0.00002 (F)   & 0.96 \\
        & 2.700 (4)   & 0.9646 (4)  &  0.00 (F)   &  0.0000  (F)   & 0.55 \\
        & 2.699 (25)  & 0.964  (4)  &  0.00 (F)   & -0.0000  (6)   & 0.68 \\  
        & 1.86  (24)  & 0.950  (F)  &  0.83 (26)  & -0.0001  (5)   & 0.68 \\
        & 2.10  (27)  & 0.955  (F)  &  0.60 (29)  & -0.0001  (6)   & 0.68 \\
        & 2.38  (30)  & 0.960  (F)  &  0.31 (33)  & -0.0001  (5)   & 0.68 \\
        & 2.75  (35)  & 0.965  (F)  & -0.05 (38)  & -0.0000  (1)   & 0.68 \\
        & 3.24  (41)  & 0.970  (F)  & -0.54 (44)  & -0.0000  (5)   & 0.68 \\
        & 3.93  (50)  & 0.975  (F)  & -1.23 (53)  &  0.0000  (5)   & 0.69 \\
        & 4.97  (64)  & 0.980  (F)  & -2.27 (66)  &  0.0000  (6)   & 0.68 \\

\hline
 4.0    & 1.230  (F)  & 0.910 (F)  &  1.26 (F)   & -0.0010 (F)    & 0.70 \\
        & 2.534  (8)  & 0.965 (1)  &  0.00 (F)   &  0.0000 (F)    & 0.87 \\
        & 2.489 (18)  & 0.956 (3)  &  0.00 (F)   & -0.0011 (4)    & 0.51 \\
        & 2.15  (17)  & 0.950 (F)  &  0.33(18)   & -0.0011 (4)    & 0.50 \\
        & 2.42  (19)  & 0.955 (F)  &  0.06(21)   & -0.0012 (4)    & 0.51 \\
        & 2.76  (21)  & 0.960 (F)  & -0.26(23)   & -0.0011 (4)    & 0.51 \\
        & 3.18  (25)  & 0.965 (F)  & -0.70(27)   & -0.0011 (4)    & 0.51 \\
        & 3.75  (30)  & 0.970 (F)  & -1.26(32)   & -0.0011 (4)    & 0.51 \\
        & 4.55  (36)  & 0.975 (F)  & -2.06(38)   & -0.0010 (4)    & 0.52 \\
        & 5.74  (45)  & 0.980 (F)  & -3.26(47)   & -0.0010 (4)    & 0.52 \\
\end{tabular}
\end{ruledtabular}
\caption{\label{tab:anomfit}
Fits to $\langle\bar \psi \psi\rangle = A m ^{1 / \delta} + B m + \langle\bar \psi 
\psi\rangle_0 $}
\end{table}

\subsection{Fits motivated by the Magnetic Equation of State}
\label{sec:EOS}
Finally we considered fits motivated by the magnetic equation of state. 
The following equation is a satisfactory parameterization
\begin{equation}
m = A \langle\bar \psi \psi\rangle + B \langle\bar \psi \psi\rangle^{\delta}\, ,
\label{eq:magneticeos}
\end{equation} 
which would of course coincide with the simple power law when A=0.
The coefficient of the linear term $A$ should vanish at a critical
point, with $A \propto (\beta - \beta_c)$. This of course explains the
smallness of A close to the transition, while
$\delta$ is the conventional magnetic exponent.
The linear term in the condensate is implied by chiral symmetry, and guarantees that 
the ratio 
\begin{equation}
\lim_{m \to 0} R_\pi 
= \frac{\partial\langle\bar\psi\psi\rangle /\partial m}{\langle\bar \psi \psi\rangle/m}
= 1
\end{equation}
approaches unity in the chiral limit and in the chirally symmetric phase.
We can view Eq.~(\ref{eq:magneticeos}) as a model for a theory
with anomalous dimensions, which incorporates the correct chiral limit.
Note that the linear term
of Eq.~(\ref{eq:magneticeos}) is of different origin than the one considered
in Eq.~(\ref{eq:anomalous}). The latter describes violations of
scaling and it is increasingly relevant at larger masses. In  Eq.~(\ref{eq:magneticeos}) 
instead, it is dominating at very small masses, away from the critical
point.

Results for this case are given in Table \ref{tab:eosfit}.
The fit $m = m(\langle\bar \psi \psi\rangle)$  
was performed with a least squares algorithm. Note that, as expected, the
significance of the linear term is very low, closer to the bulk
transition, and slightly larger by moving away from it. 
In Table \ref{tab:eosfit2} we quote the numerical solutions of the
equation $m( \langle\bar \psi \psi\rangle ) = m_{sim}$, with $m_{sim}$ the simulation 
masses, to be compared with the simulation results for the condensate.  
The agreement is very good. 

All fits clearly favor a positive value for the coefficient
of the linear term, as it should be in the chirally symmetric phase,
and within the large errors the results for the exponent are compatible
with the ones coming from the genuine power law fits. We conclude again
 in favor of chiral symmetry restoration. 
\begin{table}
\begin{ruledtabular}
\begin{tabular}{|c|c|c|c|}
$6/g_L^2 $ & A & B &  $\delta$  \\
\hline
 3.9    &  0.1(9) & 0.3(9)  & 1.1(2) \\
\hline
 4.0    &  0.3(1) & 0.077(9) & 1.3(1) \\
\end{tabular}
\end{ruledtabular}
\caption{\label{tab:eosfit}
Fits to $m = A\langle\bar \psi \psi\rangle + B\langle\bar \psi \psi\rangle^\delta $}
\end{table}
\begin{table}
\begin{ruledtabular}
\begin{tabular}{|c|c|c|c|}
$6/g_L^2$ & am & $\langle\bar \psi \psi\rangle $ & $\langle\bar \psi \psi\rangle_\mathrm{fit}$ \\
\hline
 3.9    & 0.025 & 0.07693(07) & 0.07689 \\
        & 0.040 & 0.12092(17)  & 0.12102 \\
        & 0.050 & 0.15018(10)  & 0.15010 \\
        & 0.060 & 0.17897(14)  & 0.17898 \\
        & 0.070 & 0.20768(10)  & 0.20768 \\
\hline
 4.0    & 0.025 & 0.07202(10)  & 0.07204 \\
        & 0.040 & 0.11360(10)  & 0.11355 \\
        & 0.050 & 0.14079(07) & 0.14083 \\
        & 0.060 & 0.16787(11)  & 0.16787 \\
        & 0.070 & 0.19470(13) & 0.19469 \\
\end{tabular}
\end{ruledtabular}
\caption{Comparison of the simulation results for $\langle\bar \psi \psi\rangle$ 
with the ones obtained from the fits to the Magnetic Equation of State.}
\label{tab:eosfit2}
\end{table}
\subsection{Side-by-side comparison of the two simplest scenarios}

The spirit of the analysis performed above is to see if any of
the simplest physically motivated parameterizations can account for
a condensate in the chiral limit different from zero, and we can conclude that
all analyses favor a vanishing chiral condensate. 
In this subsection we directly compare in more detail 
the genuine linear fit, Eq.~(\ref{eq:goldstone})
with $B=0$,   as this is the only fit
that produced a tiny non zero chiral condensate,
and the genuine power law fit, Eq.~(\ref{eq:anomalous}) with $B$ and 
$\langle\bar \psi \psi\rangle_0=0$, being it the simplest fit with a
 $\chi^2$ in an acceptable statistical range. In the rest of this section 
we refer to these
fits as 'linear' and 'power-law', respectively. 

\begin{figure}
\vspace*{0.5 truecm}
\includegraphics[width=8.0truecm]{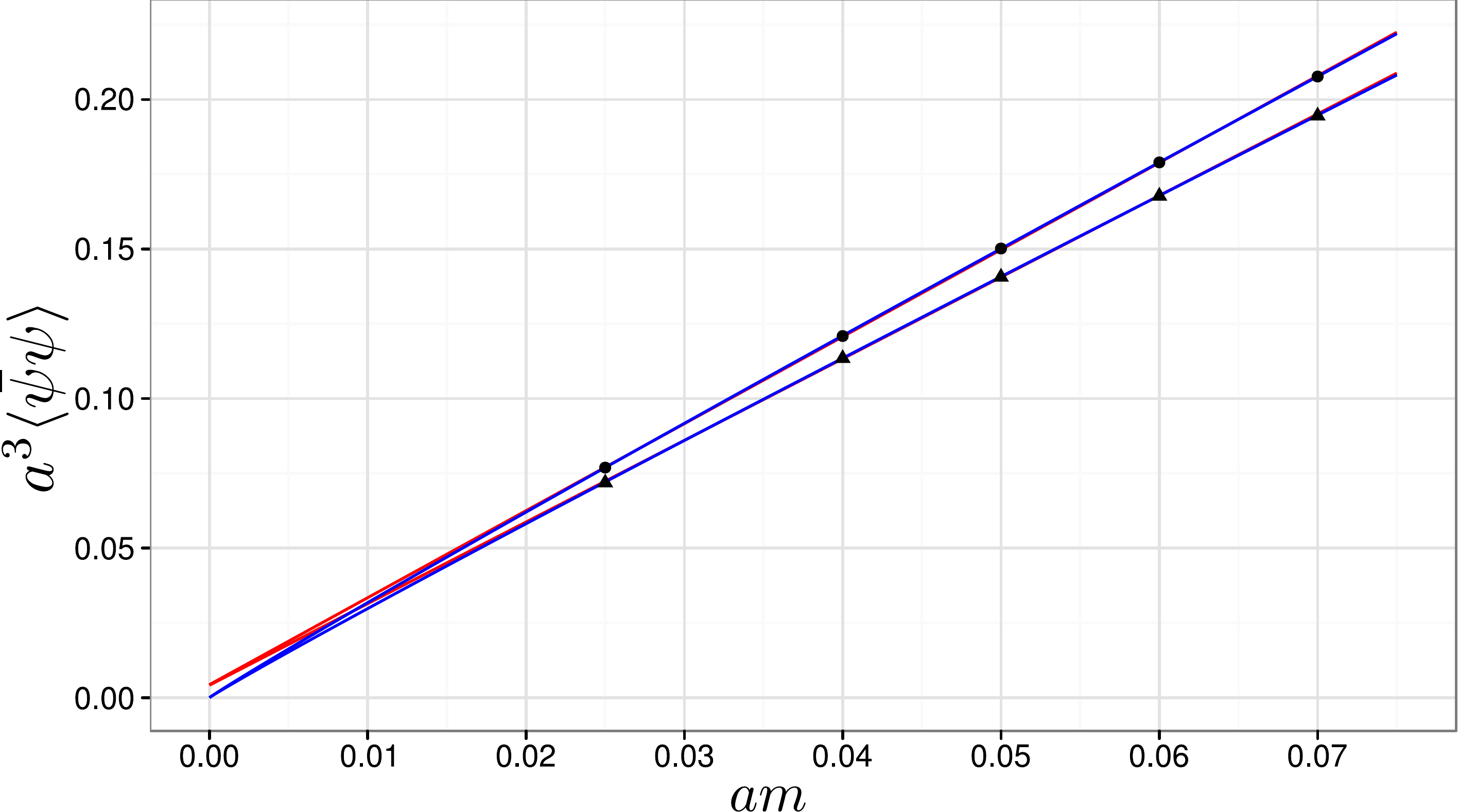}
\caption{\label{fig:condensate_simple} {(color  online) Fits to  the chiral
condensate   measured   at   $6/g_L^2=3.9$  (circles)   and   $6/g_L^2=4.0$
(triangles), with linear fits  shown in
red and  power-law fits drawn in blue.}}
\end{figure}

The measured  values of  the chiral condensate  and those predicted  by the
linear  and   power-law  fits  are  shown   in  Table~\ref{tab:pbp}.
In Fig.~\ref{fig:condensate_simple}
 the  measured data with  superimposed fits
are shown.  Of course, since the range of variability of the chiral condensate
is exceedingly larger than its errors, it is impossible to
appreciate by eye the quality of the fits on this scale. 
\begin{table}
\begin{ruledtabular}
\begin{tabular}{|c|c|c|c|c|}
  $6/g_L^2$ &  $am$ & $a^3\langle\bar{\psi}\psi\rangle_\mathrm{measured}$ &
$a^3\langle\bar{\psi}\psi\rangle_\mathrm{linear}$                         &
$a^3\langle\bar{\psi}\psi\rangle_\mathrm{power}$  \\ \hline  3.9 &  0.025 &
0.07693(07)  & 0.07705  & 0.07692  \\  & 0.040  & 0.12092(17)  & 0.12069  &
0.12105  \\  & 0.050  &  0.15018(10)  & 0.14978  &  0.15013  \\  & 0.060  &
0.17897(14)  & 0.17887  & 0.17899  \\  & 0.070  & 0.20768(10)  & 0.20796  &
0.20769 \\ \hline 4.0 & 0.025 &  0.07202(10) & 0.07237 & 0.07212 \\ & 0.040
& 0.11360(10)  & 0.11331  & 0.11350 \\  & 0.050  & 0.14079(07) &  0.14060 &
0.14077  \\  & 0.060  &  0.16787(11)  & 0.16789  &  0.16785  \\  & 0.070  &
0.19470(13) & 0.19518 & 0.19477 \\
\end{tabular}
\end{ruledtabular}
\caption{\label{tab:pbp}Measurements  of  the  chiral  condensate  at  $N_s
\times N_t = 24^3\times 24$ for  two values of the coupling $6/g^2=3.9$ and
$4.0$, and  a range  of bare  quark masses $am$,  together with  the values
predicted by the fits to a linear and a power-law model.}
\end{table} 
\begin{figure}
\vspace*{0.5 truecm}
\includegraphics[width=8 truecm]{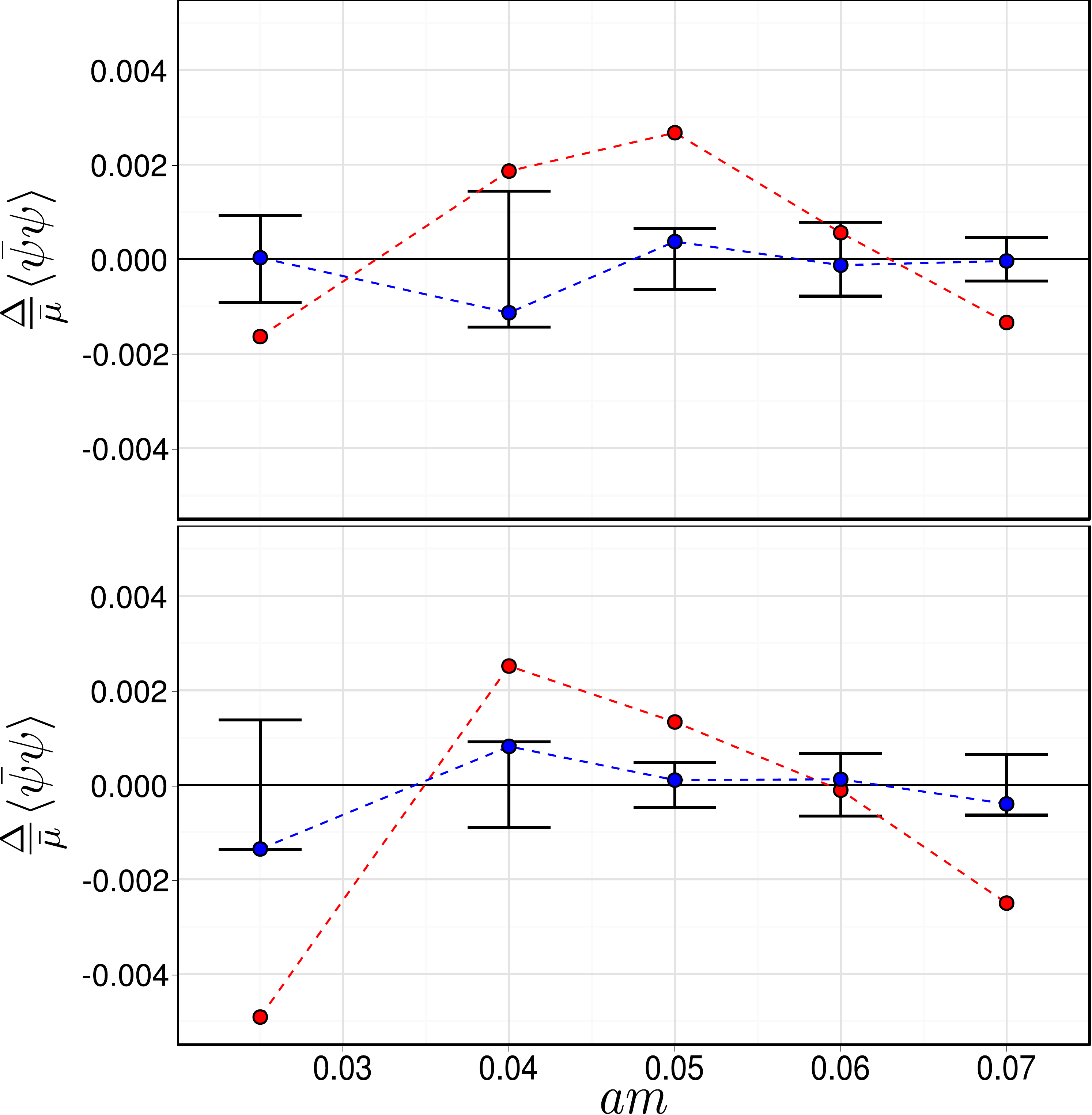}
\caption{\label{fig:condensate_rel} {(color  online) Deviations $\Delta$ of
the  fitted  value  from  the  measured  value  $\bar\mu$  for  the  chiral
condensate, rescaled by the measured value itself. Error bars represent the
relative  (rescaled  by  the  data)  standard  deviation  on  the  measured
value. Deviations  from the  prediction with  a fit to  the linear  form 
are   given  in  red,   while  those  for  a   fit  to
a power-law are  given in blue.  The top graph displays  results for
$6/g_L^2=3.9$, the lower graph those  for $6/g_L^2=4.0$. The linear form 
 shows tension  with the  data for  both values  of the
coupling, which is quantitatively seen in the larger $\chi^2$ value.}}
\end{figure}
\begin{figure}
\vspace*{0.5 truecm}
\includegraphics[width=8.0truecm]{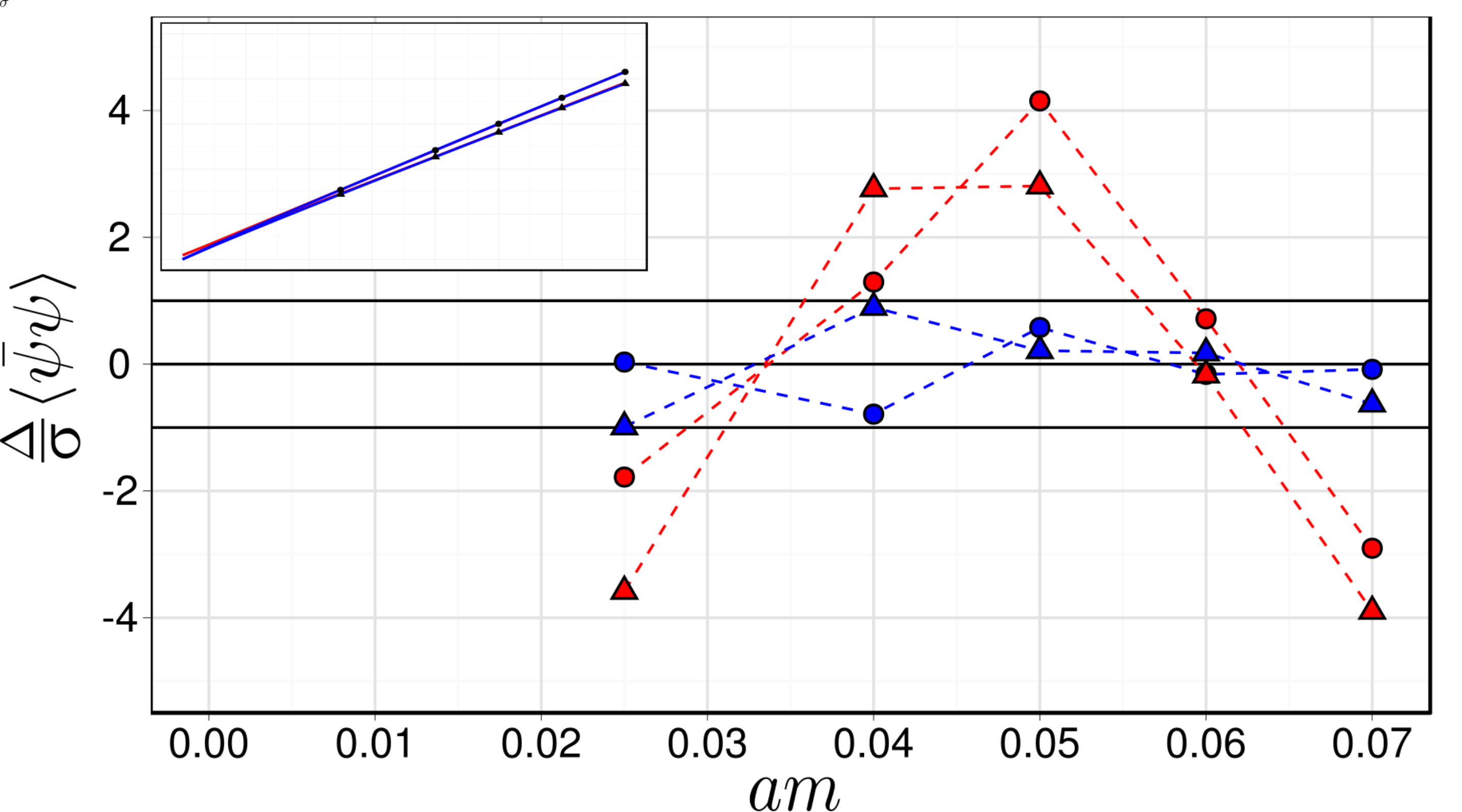}
\caption{\label{fig:condensate} {(color online)  Deviations $\Delta$ of the
fitted value from  the measured one for the  chiral condensate, rescaled by
the standard deviation $\sigma$ of  each measurement. Results are shown for
both $6/g_L^2=3.9$  (circles) and  $6/g_L^2=4.0$ (triangles), with  fits to
the linear form   shown in red and  fits to the power
law  drawn in  blue. The data and  corresponding fits
are displayed on linear scales in the inset .}}
\end{figure}
A more effective description of the relative quality of the fits is offered by
Fig.~\ref{fig:condensate_rel}   and  Fig.~\ref{fig:condensate}.  
 In Fig.~\ref{fig:condensate_rel} we plot the difference between
the chiral  condensate predicted by the  fits and the data,  divided by the
data themselves.  The tension between  fitted and numerical results for the
linear form is quite evident.  The  pattern of the deviations in the linear
fits indicates a  significant curvature, which is reflected  in the quality
of the fit.  The pattern of the  residuals of the power law  fit is instead
far   less   structured   and   statistically   insignificant   throughout.
Fig.~\ref{fig:condensate}  offers in  our  opinion the  most  clear way  of
visualizing  the   deviations  by  plotting  the  same   difference  as  in
Fig.~\ref{fig:condensate_rel},  this time  divided by  the  error $\sigma$.
The horizontal band  indicates the boundary of one  standard deviation, and
the points obtained  by a power law fit nicely fall  within it, while again
the tension with the linear form appears.  These results thus 
confirm a
strong  preference for  the  restoration  of chiral  symmetry  at the  weak
coupling  side of the  transition, as  was inferred  from sections \ref{sec:systematic}
to \ref{sec:EOS}.

It is  clear that additional data  at even lighter masses  will improve
the   discriminating  power  of   these  fits   and  eventually   allow  to
significantly constrain  the linear contributions.   The presence of
curvature  in the  data and  the very  good quality  of the  power-law fit,
having barred finite volume effects, is also an indication that 
we are not in the heavy quark limit. In addition, one
could  also study  the  analogous of  the  GMOR relation  of broken  chiral
symmetry, and variations  of it in terms of the scalar  meson mass, by also
measuring  the pion  decay constant  $f_\pi$ in  the chiral  limit  and the
scalar mass.

\section{Spectrum analysis}
\label{sec:spectrum}

An alternative approach to the study of the symmetry of a phase is
offered by the spectrum analysis \cite{Kocic:1992is}.
Particularly useful quantities for this type of study are the masses of the
ground  state excitations  in the  pseudoscalar and  vector  channels, with
slight abuse of  nomenclature from QCD referred to as  the $\pi$ and $\rho$
masses.  We defer  to  future  work the  exploration  of other  interesting
observables, such  as the  ratio of the  scalar and pseudoscalar  masses or
equivalently   the   ratio    of   transverse   and   longitudinal   chiral
susceptibilities.

\subsection{Spectrum and Chiral Symmetry}
\begin{figure}
\vspace*{0.5 truecm}
\includegraphics[width=8.0truecm]{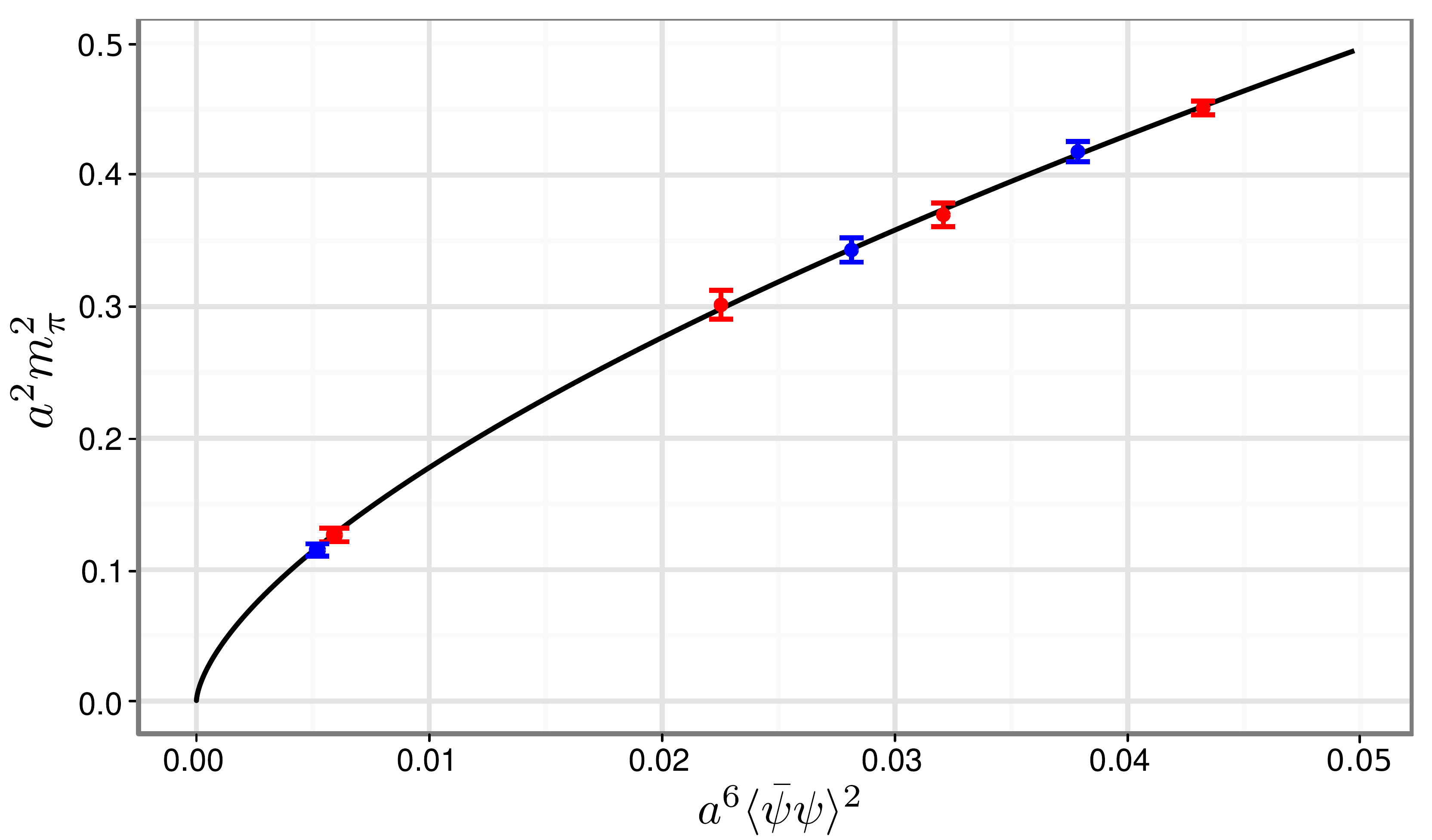}
\caption{\label{fig:pion_pbp_plot} {(color online) The relation between the
chiral condensate  and the pion  mass, for $6/g_L^2$  = 3.9 (blue)  and 4.0
(red). The  line represents a power law fit  to the combined  data, 
the results of  which are
reproduced in Table~\ref{tab:pion_pbp_fit}.}}
\end{figure}

A powerful way to distinguish  between symmetric and broken chiral symmetry
\cite{Kocic:1992is} is  to plot  the pseudoscalar mass  as a function  of the
chiral condensate, as in Fig.  \ref{fig:pion_pbp_plot}.  We have considered
the same range of bare  fermion masses used in section \ref{sec:chiral} for
the chiral extrapolation of the condensate.   The data are best fitted by a
simple   power-law   form  and   the   results   are   reported  in  Table
\ref{tab:pion_pbp_fit}.   They  clearly  suggest  that chiral  symmetry  is
restored and  that the  theory has anomalous  dimensions. In  the symmetric
phase and  in mean field  \cite{Kocic:1992is}, we expect a  linear dependence
with  non  negative intercept.  The  presence  of  anomalous dimensions  is
responsible  for negative curvature  - noticeably  opposite to  what finite
volume effects would  induce - and a zero intercept. The  same graph in the
broken  phase would  show the  opposite curvature  and extrapolate  with a
negative intercept.

\begin{figure}
\vspace*{0.5 truecm}
\includegraphics[width=8.0truecm]{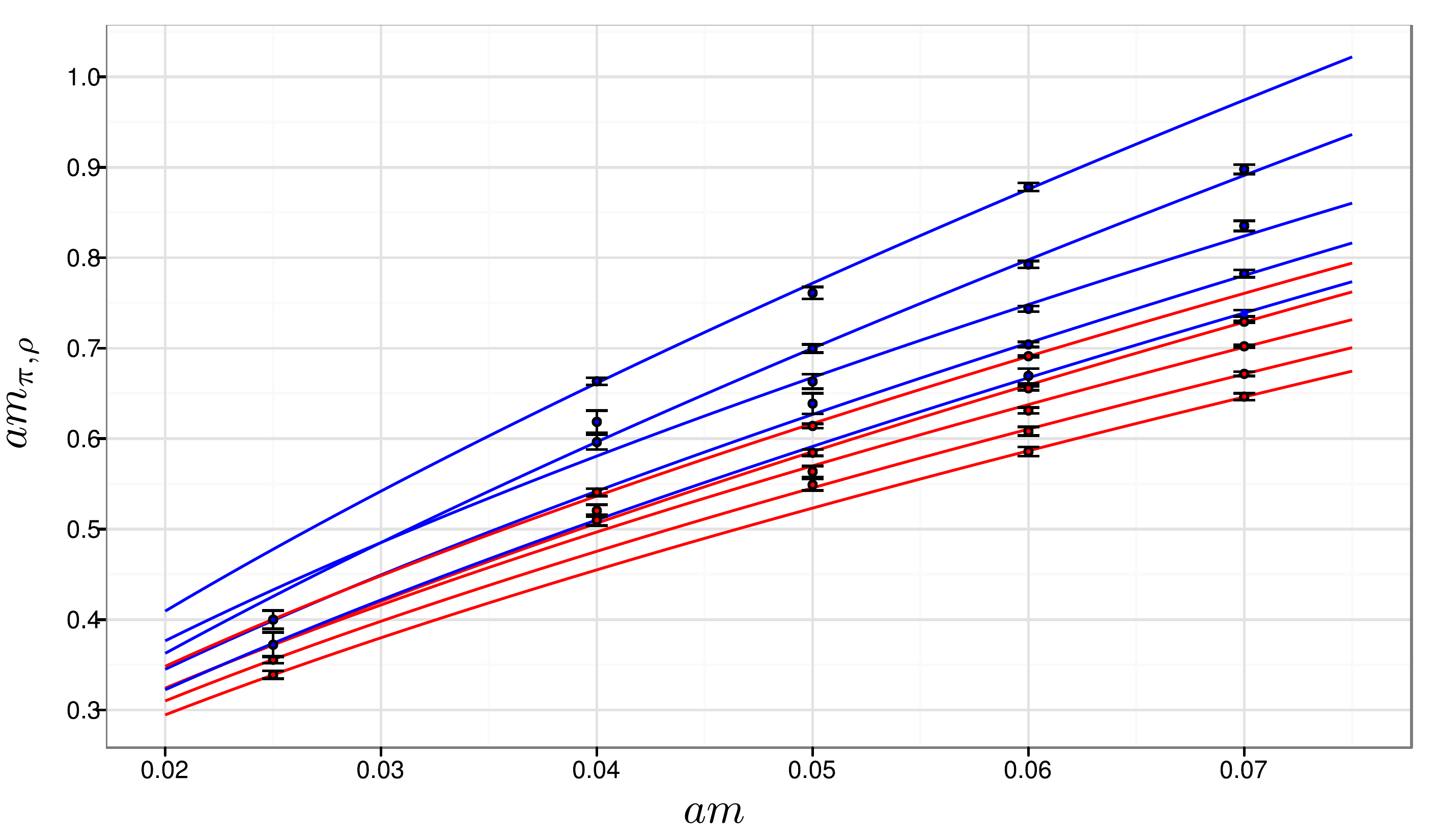}
\caption{\label{fig:meson_mass_plot}  {(color online) The  relation between
the bare quark mass and the masses  of the pion (red) and rho meson (blue),
for  $6/g_L^2$  = 3.6,  3.7,  3.8,  3.9 and  4.0  from  the uppermost  line
down. Power law fits to the separate values of beta are provided.}}
\end{figure}

This result gives  also further confidence that the  fermion masses used in
this study  are not too light, so  that they do not  significantly feel the
finite volume,  and not  too heavy, so  that they  are not blind  to chiral
symmetry.
\begin{table}
\begin{ruledtabular}
\begin{tabular}{|c|c|c|}
  $6/g_L^2$  & parameter           & value    \\ 
\hline  
        3.9  & $A$                 & 3.350(70)  \\
             & $\delta_\chi$       & 0.639(6) \\ 
             & $\sqrt{\chi^2\,\mathrm{dof}}$ & 0.73     \\ 
\hline 
         4.0 & $A$                 & 3.500(40) \\ 
             & $\delta_\chi$       & 0.649(3) \\
             &$\sqrt{\chi^2\,\mathrm{dof}}$ & 0.46 \\
\hline
 combined    & $A$                 & 3.400(50) \\ 
             & $\delta_\chi$       & 0.642(4) \\ 
             &$\sqrt{\chi^2\,\mathrm{dof}}$ & 0.70 \\
\end{tabular}
\end{ruledtabular}
\caption{\label{tab:pion_pbp_fit}  Results of fits  to the  functional form
$(a m_\pi)^2 = A (a^3 \langle\bar\psi\psi\rangle)^{2\delta_\chi}$. Fits are
performed to the separate values  of the coupling constant and the combined
data set.}
\end{table} 
In  Fig.~\ref{fig:meson_mass_plot}  we report  on  the  measured values  of
$m_\pi$  and $m_\rho$  as a  function of  the bare  fermion mass.  Here the
lightest point  at $am=0.025$ for the  vector mass is  absent, but a curvature
can still be appreciated.  
Simulations were done on $16^3\times 24$ volumes,
while a  set of  measurements at larger  volumes showed that  finite volume
effects were under control. 
The  mass dependence
shown in Fig.~\ref{fig:meson_mass_plot} hints  again at a few properties of
a chirally symmetric phase. We have fitted both the pion and the
rho mass to a power law
\begin{equation}
m_{\pi,\rho} = A_{\pi,\rho} m ^{\epsilon_{\pi,\rho}}
\end{equation}
with the results
$A_\pi = 3.41(21)$, $\epsilon_\pi = 0.61(2)$,  
$A_\rho = 4.47(61)$, $\epsilon_\rho = 0.66(5)$  at $6/g_L^2 = 3.9$, and 
$A_\pi = 3.41(21)$, $\epsilon_\pi = 0.61(2)$,
$A_\rho =  4.29(11)$, $\epsilon_\rho  = 0.66(1)$ at $6/g_L^2 =  4.0$. 
The accuracies  of these
fits
are  not  comparable  with  those  achieved  by  the  fits  to  the  chiral
condensate,
however they allow to draw a few conclusions.
 First, the mass dependence  of the vector and
pseudoscalar mesons is well fitted by a power-law.  Second, it is 
also relevant
that  the exponents are  not unity  and $\epsilon_\pi  \neq 1/2$.   The  latter result
immediately tells  that the pion  seen here is  not a Goldstone boson  of a
broken chiral symmetry.  In addition, both mesons have  masses scaling with
roughly the  same power,  as it should  be in  a symmetric phase,  and with
increasing degeneracy towards the chiral  limit.  The exponent of the power
law  being  not  one,  confirms  that   we  are  not  in  the  heavy  quark
regime. 

These results are confirmed in a more visual way by looking at the behavior
of the mass ratio. 
Fig.~\ref{fig:Ratio} shows  the ratios of measured  pseudoscalar and vector
masses, for a fixed  coupling and as a function of the  bare quark mass. We
have superimposed  the ratios  of the best  fits to  the raw mass  data, as
explained in section \ref{sec:betafunc}.   It is immediately clear that the
ratio increases as  the quark mass approaches zero,  a behavior opposite to
what is expected for a Goldstone pion.  
%
Notice also that the mass ratio  should be one for exact conformal symmetry
in the  chiral limit: we do not yet  observe that,
since, as  explained in  Section \ref{sec:strategy}, conformal  symmetry is
expected to be broken by Coulombic  forces in the region of parameter space
probed  by this  study.  On the  other  hand, the  trend  towards unity  as
decreasing the lattice coupling $g_L$ is evident, and certainly worth further exploration. 
See also Ref.~\cite{DelDebbio:2009fd} for a study of the spectrum. 
\subsection{Spectrum, lattice spacing and the beta function}
\label{sec:betafunc}
\begin{figure}[t]
\vspace*{0.5 truecm}
\subfigure[Interpolations    of   meson   masses.]{\label{fig:Spectral_fit}
\includegraphics[width=8.0truecm]{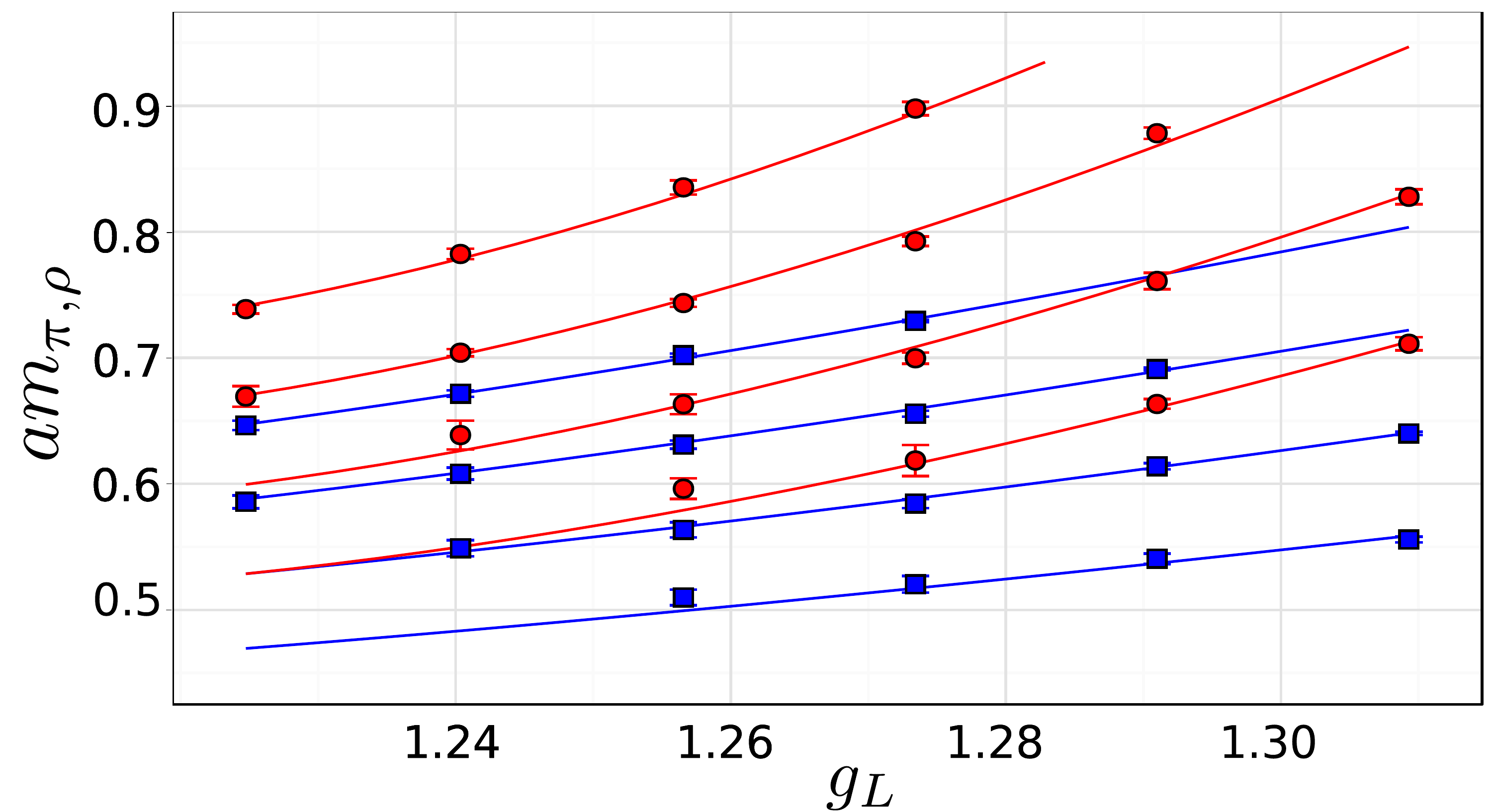}}   \subfigure[$\pi$  to
$\rho$    mass   ratio    with   fits    superimposed   ]{\label{fig:Ratio}
\includegraphics[width=8.0truecm]{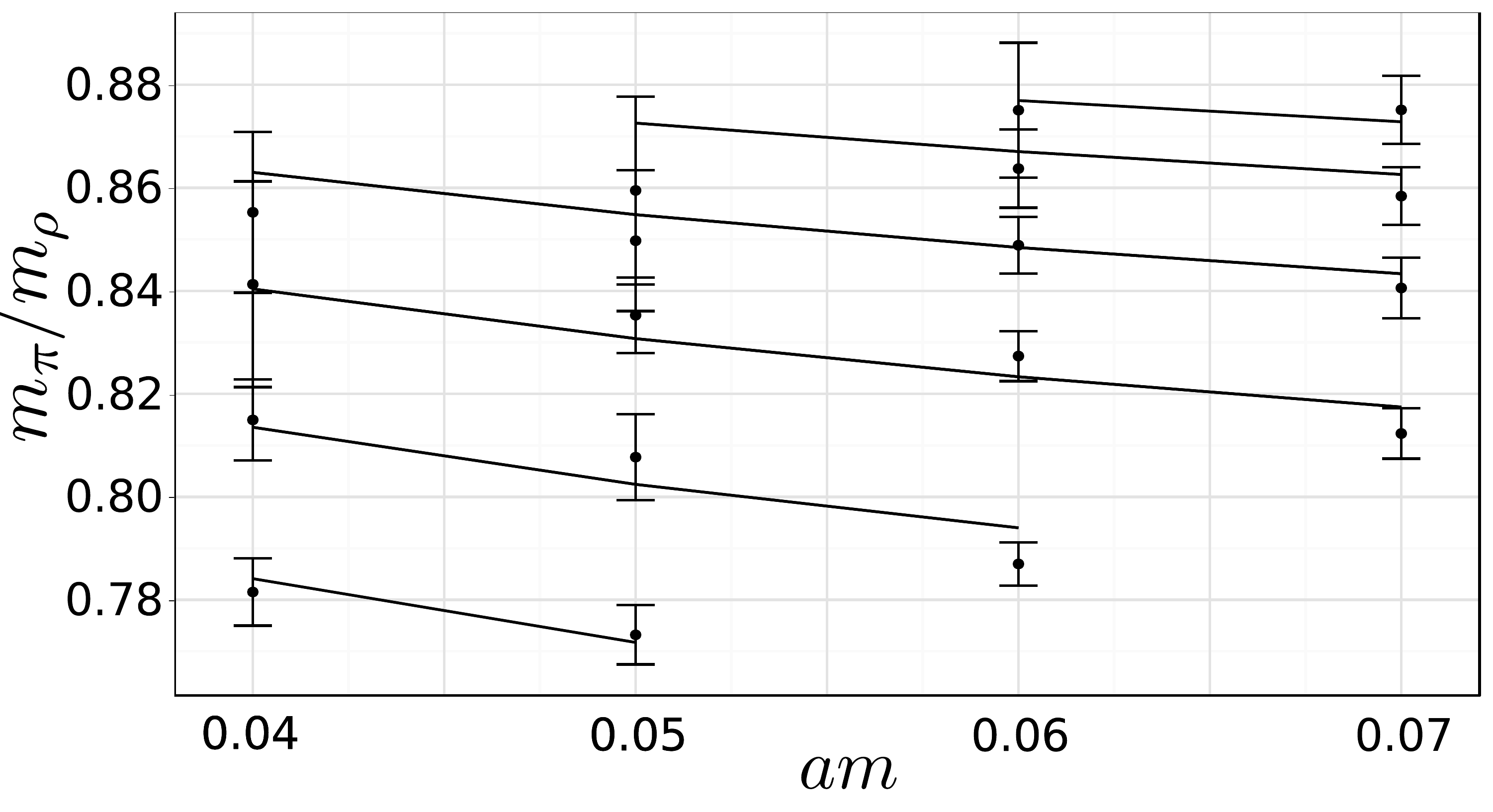}}
\caption{(color  online) (a)  Measurements of  the pseudoscalar  (blue) and
vector (red) masses  versus lattice coupling at several  values of the bare
quark  mass, from  bottom  to top  $am=0.04$,  0.05, 0.06  and 0.07.  Lines
displayed  represent   a  global  parameterization,  with   a  mixed  $O(m)$
polynomial  quark mass dependence  and $O(\beta^2)$  polynomial dependence,
with lattice parameter $\beta =  6/g_L^2$, and producing a reduced $\chi^2$
per  d.o.f. just  over  unity  for both  channels.  Errors include  fitting
systematics  from combining  several  methods. (b)  The  measured $\pi$  to
$\rho$ mass  ratio as a function  of the bare mass  and decreasing coupling
$g_L$,  bottom to  top $6/g_L^2=3.5$  to  $4$. The  superimposed lines  are
ratios of the best fits in Fig.~\ref{fig:Spectral_fit}.}
\end{figure}
\begin{figure}
\vspace*{0.5 truecm}
\includegraphics[width=8.0truecm]{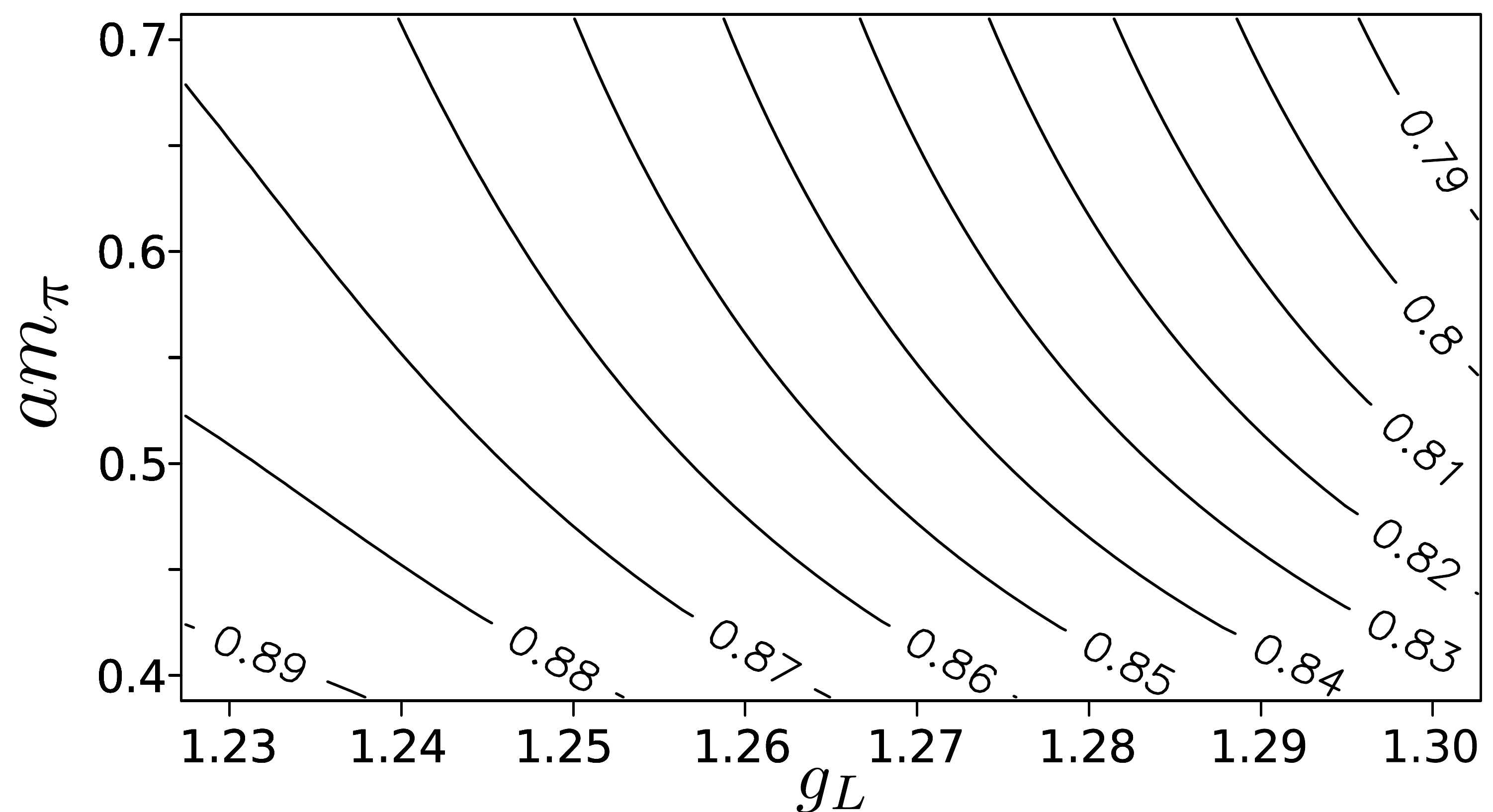}
\caption{\label{fig:PionMass_slope}  {Pseudoscalar   mass  along  lines  of
constant  physics. The  physical  pion  mass is  identical  along lines  of
constant physics,  such that  a decreasing pion  mass in lattice  units for
increasing $g_L$,  implies an  increase of the  lattice spacing  for weaker
couplings,  signature of  the Coulomb-like  phase. The  pseudoscalar  mass along
lines  of  constant physics  was  constructed  by  interpolating along  the
isolines of  the ratio of  the separate interpolations of  pseudoscalar and
vector  masses.  Different  polynomial interpolations  produced  compatible
results,  and  in  agreement   with  a  non-interpolated  analysis  of  raw
data.  Labels  give  the  value  of  the ratio  $m_\pi  /m\rho$  along  the
isolines. }}
\end{figure}
We used the spectrum results to  determine the lines of 
``constant physics'' in
the two dimensional parameter space $g_L$  and $am$, the bare quark mass of
degenerate fermions, following the same strategy which was successful 
for $N_f = 16$~\cite{damgaard_lattice_1997}. 
Along these lines the coupling and masses are all
functions  of the lattice  spacing $a$.  Since all  dimensionful quantities
measured on the  lattice will be expressed in terms  of the lattice spacing
and  will therefore  vary with  $g_L$  even if  they do  not physically,  a
dimensionless quantity has to be  taken as a reference. A convenient choice
is the ratio of the $\pi$ and $\rho$ masses.
Before continuing, let us specify that the same caveat
as in Ref.\cite{damgaard_lattice_1997} applies: 
since we are at strong coupling, there is no
guarantee that the system can be described in terms of a one-parameter
beta function. This implies, for instance, that the lines of
``constant physics'' determined by use of certain observables might not
match those determined using other observables.
If multiple bare couplings are needed, it might happen that 
the change of physics produced by changing
only one bare coupling will not be compensated by a change of mass.
So our lines of ``constant
physics'' are, strictly speaking, 
lines of constant $m_\pi/m_\rho$ ratio. We will
show in the following that in order to keep this ratio constant the 
bare parameters $am $ and $g_L$ controlling
the simulations should be tuned as if we had a one parameter, positive beta 
function. 

In Fig.~\ref{fig:Spectral_fit} we report  on the measured values of $m_\pi$
and $m_\rho$ as a function of the bare coupling, while Fig.~\ref{fig:Ratio}
shows the  ratios of measured pseudoscalar  and vector masses,  for a fixed
coupling and as a function of the bare quark mass. We have superimposed the
ratios of the  best fits to the raw mass data,  confirming the good quality
of the  interpolations derived  in Fig.~\ref{fig:Spectral_fit} and  used to
produce Fig.~\ref{fig:PionMass_slope}.

It is immediately evident from  Fig.~\ref{fig:Ratio} that, in order to keep
the ratio constant, we  should simultaneously decrease the lattice coupling
$g_L$ and increase  the bare mass. These results  already indicate that the
lattice spacing increases  while decreasing the coupling. This  is the same
behavior  as observed  for $N_f=16$,  and  of the  pion to  sigma ratio  in
QED. It is  also expected of a one parameter beta  function with a positive
sign.

To  refine the  analysis, and  express the  result in  terms of  a physical
observable,  we  proceed as  follows.  Given  the  ratio $m_\pi/m_\rho$  at
reference values  of $g_L$ and $a  m$, one can determine  a value $a^\prime
m^\prime$  at  coupling $g_L^\prime$  in  the  surroundings  of $g_L$  that
reproduces  the same  ratio and  thus  lies on  the same  line of  constant
physics. This is implemented by  fitting the measured values of both masses
to  a  parameterization,   as  shown  in  Fig.~\ref{fig:Spectral_fit},  then
determining the  isolines from the  ratio of the two  parameterizations. The
physical pseudoscalar  mass being constant  along each of these  lines, the
ratio of measured values $a m_{\pi} / a^\prime m_{\pi}$ directly determines
the ratio of the lattice spacings $a / a^\prime$. If a decrease in $g_L$ is
associated with  an increase in the  lattice spacing, the sign  of the beta
function   is    positive,   i.e.   that   of   the    Coulomb-like   phase   in
Fig.~\ref{fig:manyphases}.

Generalizing, along the lines of ``constant  physics'' the slope of the line of
measured values of the pseudoscalar mass is a direct measure of the sign of
the beta function.  Fig.~\ref{fig:PionMass_slope} provides evidence for the
Coulomb-like phase, with a positive sign of the beta function, in full agreement
with  the more  naive discussion  of Fig.~\ref{fig:Ratio}.  Since  the beta
function  is known  to  be negative  in  the continuum  limit, our  results
indicate a zero of the beta  function at some intermediate coupling $g$. We
emphasize that the  location of this zero is  regularization dependent,
and we reiterate the caveat at the beginning of this section. 
Further, 
 we do  not claim to have directly studied the  physics around the IRFP
itself.  The latter  type  of  study is  notoriously  difficult, while  the
strategy presented here aims at probing the emergence of conformality in an
indirect way.
\section{\label{sec:conclusion}Summary and Outlook}

We summarize here the main findings of our study:

\noindent i)  For an  SU(3)  gauge  theory with  three  unrooted staggered  fermions,
corresponding to twelve continuum flavors,  we have observed a lattice bulk
transition or crossover which is clearly of a non-thermal nature.

\noindent ii) We  have studied  the realization  of the chiral  symmetry on  the weak
coupling  side of  this transition:  the  analysis of  the order  parameter
favors chiral symmetry restoration.

\noindent iii)  A study  of the  spectrum in  the weak  coupling phase  close  to the
transition favors chiral symmetry restoration as well.

\noindent iv) We have  derived the lines of ``constant physics''  and inferred a positive
sign of the beta function, again implying the emergence of a Coulomb-like phase.

The above  results provide evidence  towards the existence of  a symmetric,
Coulomb-like  phase  on  the  weak   coupling  side  of  the  lattice  bulk
transition.     In    the    scenario    of    Refs.~\cite{appelquist_1996,
miransky_conformal_1997}  and  Fig.~\ref{fig:manyphases},  such  a  Coulomb-like
region must be entangled to the presence of a conformal infrared fixed point
for  the  theory  with   twelve  continuum  flavors,  without  any  further
transition  at weaker coupling.  Such a  Coulomb-like phase  is not  expected in
ordinary QCD.  We  reiterate that the evidence provided  is indirect, while
we do not address the physics at the infrared fixed point.

A few directions  are a natural extension of this  work. An accurate chiral
extrapolation of the chiral condensate  in the strong coupling phase, would
allow to determine the precise  location of the chiral phase transition (or
crossover). Establishing  the nature of  such a bulk transition  might shed
light  on the  possible  emergence of  an  ultraviolet fixed  point in  the
continuum  theory  at  strong  coupling \cite{Kaplan:2009kr}.  It  is  also
important  to notice that  a way  to discriminate  between the  scenario of
Refs.~\cite{appelquist_1996,    miransky_conformal_1997}   and    the   one
originally proposed  in Ref.~\cite{banks_phase_1982}  is the presence  of a
chiral transition  towards a broken  phase at weaker couplings.  While both
scenarios share the  presence of conformality and of  a Coulomb-like phase, only
in the  first a range of theories  exists -- the conformal  window -- where
confinement and chiral symmetry breaking do not occur at weak coupling. For
a recent review on the subject, see Ref. \cite{Pallante2009}.  In addition,
more extended  results on the mass  spectrum, in particular  an analysis of
the  chiral partners, would  shed further  light on  the pattern  of chiral
symmetry breaking and restoration for this theory. Work in these directions
is  in progress.  Alternative  studies based  on the  Renormalization Group
analysis as  proposed in \cite{DeGrand:2009mt} will  provide an independent
and  valuable  tool to  investigate  these  systems.  Such studies  aim  to
directly  probe the  existence of  an infrared  fixed point  and complement
indirect  searches for conformal  behavior in  $SU(N)$ gauge  theories with
matter content.

\begin{acknowledgments}
This  work was  in  part based  on  the MILC  public  lattice gauge  theory
code. We  thank M. Bochicchio, F. Di  Renzo, J. Kuti, F.  Sannino, C. deTar
for comments and discussions. Computer  time was provided through the Dutch
National Computing Foundation (NCF) and the University of Groningen.
\end{acknowledgments}

\bibliographystyle{apsrev} 
\bibliography{Conformal}

\end{document}